# Cascade of fractional quantum Hall states in 2D system

Zhimou Chen[1], Jiaojie Yan[2], Yuxuan Zhu[1], Zhe Cui[1], Loren N. Pfeiffer[3], Kenneth W. West[3], Kirk W. Baldwin[3], Adbhut Gupta[3], Yang Liu[1], Wei Zhu[4,5], Wenchen Luo[6], Ying-Hai Wu[7,*], Shuai Yuan[8,*] and Xi Lin[1,9,10,*]

[1]International Center for Quantum Materials, Peking University, Beijing 100871, China
[2]Max Planck Institute for Solid State Research, Stuttgart 70569, Germany
[3]Department of Electrical Engineering, Princeton University, Princeton 08544, USA
[4]Institute of Natural Sciences, Westlake Institute of Advanced Study, Hangzhou, 030024, China
[5]School of Science, Westlake University, Hangzhou, 030024, China
[6]School of Physics, Central South University, Changsha 410083, China
[7]School of Physics and Wuhan National High Magnetic Field Center, Huazhong University of Science and Technology, Wuhan 430074, China
[8]Department of Physics, University of Washington, Seattle 98195, USA
[9]Hefei National Laboratory, Hefei 230088, China
[10]Interdisciplinary Institute of Light-Element Quantum Materials and Research Center for Light-Element Advanced Materials, Peking University, Beijing 100871, China

*Corresponding authors
E-mails: yinghaiwu88@hust.edu.cn; shuaiy81@uw.edu; xilin@pku.edu.cn

## Abstract

The observation of the fractional quantum Hall (FQH) effect in 2D electron gases ushered in investigations of topological phases driven by strong electron correlations. Their remarkable features include fractionalized elementary excitations, gapless boundary states, and non-trivial quantum entanglement patterns. Thanks to persistent efforts in the building of new platforms and making higher- quality samples, a diverse plethora of FQH states have been unveiled in experiments. We report a systematic study of ultrahigh-quality GaAs/AlGaAs quantum wells with mobility up to $3.7\times10^7\ \mathrm{cm^2V^{-1}s^{-1}}$ using quantum transport measurements in nuclear adiabatic demagnetization and dilution refrigerators down to 1 mK. In addition to many FQH states that have already been identified in previous work, new longitudinal resistance dips are observed at filling factors 17/33 and 15/31. The application of an in-plane magnetic field causes disparate variations of the FQH states. The theoretical foundation of these states is discussed in the framework of composite fermion theory. While most fractions can be explained as non-interacting composite fermions forming integer quantum Hall states, a few states correspond to FQH states of composite fermions that arise from residual interaction between them. We summarize the observed fractions in the range of $0 < \nu < 2$ and propose a pattern to account for their experimental appearance that provides an intuitive picture about the relative strengths of different FQH states.

## Introduction

Technological advances since the 1960s have made it possible to confine electrons in two dimensions, which paved the way for the observation of the quantum Hall effect in high magnetic fields[1,2]. In low-temperature electrical transport measurements, the Hall resistance exhibits plateaus at certain quantized values whereas the longitudinal resistance is exponentially suppressed. If the Hall resistance of a state is an integer (fractional) multiple of $h/e^2$, it is called an integer (fractional) quantum Hall [IQH (FQH)] state. The integer states can be understood as multiple Landau levels (LLs) fully occupied by free electrons such that an energy gap appears in the bulk of a system. For the fractional states, at least one LL is partially occupied and there is no energy gap due to single-particle physics, so the Coulomb interaction must be invoked to generate many-body energy gaps. It is generally true that interaction-induced gaps are substantially smaller than cyclotron gaps. The magnetic field range in which an FQH state appears could be quite narrow. While the Hall resistances of some IQH states are precisely quantized so they can serve as a resistance standard[3,4], the measured resistance for most FQH states does not reach the same level of accuracy. As for the longitudinal resistance, minimal values observed in IQH states are also considerably smaller than those of FQH states. From an experimental viewpoint, a local minimum in the longitudinal resistance is usually taken as a promising hint about the presence of an FQH state or other gapped phases.

In theoretical treatments of FQH states, a widely adopted approximation is to restrict electrons to one LL (possibly with other discrete degrees of freedom such as spin). Since the kinetic energy is an inconsequential constant, the physics is fully determined by interaction between electrons. This problem is very challenging because perturbative techniques are not useful, but Laughlin found a surprisingly simple real space wave function that can explain the first observed FQH state at filling factor $\nu = 1/3$, and revealed that its elementary excitations carry $e/3$ charge[5]. It was demonstrated in subsequent works that these excitations obey fractional braid statistics[6,7], thus making the abstract idea of anyons a physical reality[8–10]. Experimental investigations quickly uncovered FQH states at various other filling factors and most of them have odd denominators. This fact was rationalized by the composite fermion (CF) theory[11]. It not only provides an intuitive picture for many aspects of FQH physics but also enables quantitative calculations whose results compare favorably with experiments[12]. The existence of fractionalized excitations is a common feature of FQH states. Experimental measurements of fractional charge were reported in 1997[13,14], but compelling evidence for fractional braid statistics was only gathered in the past few years[15–19].

Electron-doped GaAs quantum wells have long served as the primary system for studying FQH states, whose mobility has reached the order of $10^7\ \mathrm{cm^2V^{-1}s^{-1}}$ for years[20]. Many other systems, including hole doped GaAs, AlAs, ZnO, graphene, black phosphorene, and transition metal dichalcogenides (TMDs) have also been demonstrated as versatile platforms despite their lower mobilities[21–38]. Furthermore, FQH states in the absence of an external magnetic field, namely fractional quantum anomalous Hall states or fractional Chern insulators, have been observed in twisted $MoTe_2$ and rhombohedral multilayer graphene[39–44]. In all these systems, the majority of the observed fractions also fit into the CF framework. The Laughlin-type states could have energy gaps as large as about 10 K[45], but other FQH states are generally more fragile. When the energy gap of a state is small, the best possible evidence supporting its existence may be a longitudinal resistance dip, yet this kind of identification could be misleading. In fact, other many-body states such as bubble states can also produce a vanishing longitudinal resistance or a dip, and the competition between two peaks (for trivial or non-trivial reasons) may also create a superficial dip. Even if a dip is accompanied by a quantized Hall plateau, there could still be other origins. For example, partial transmission and reflection of edges at a density junction give rise to accurately quantized Hall resistance at 3/2, 9/4, 17/11, and 16/13[46,47].

Given the disparate properties of these platforms and the subtleties in experimental data analysis, a comprehensive summary of existing experimental results from state-of-the-art samples is highly desirable. If a pattern can be established for the known FQH states, and features that are consistent with this pattern are spotted in upcoming devices, it would be more convincing to claim the existence of new FQH states. Indeed, the states that find simple explanations within the CF theory could serve as a backbone of this strategy. On the other hand, a violation of the pattern might imply novel physics and calls for more in-depth studies. This is the case for the second LL of electron-doped GaAs, where a prominent even-denominator state is observed at $\nu$ = 5/2[48] and the number of odd-denominator states is much smaller than in the lowest LL[49,50]. The 5/2 state is generally believed to be of the Moore-Read type[51] in which elementary charged excitations exhibit non-Abelian braid statistics[52,53]. In GaAs wide quantum wells and bilayer graphene (BLG), coexistence of many CF states and one half-filled state has been observed[54,55].

## Results

### *Polar coordinate representation of FQH states*

An expanded view of the longitudinal resistance in the vicinity of $\nu$ = 1/2 is presented in Fig. 1(d). The observed FQH states are organized using a polar system in Fig. 2(a). To make a clear presentation, the fractional part $\nu_\mathrm{f}$ of a filling factor $\nu$ is used. These two variables are the same when $\nu$ < 1, and we define $\nu_\mathrm{f} = \nu - 1$ or $2 - \nu$ when $1 < \nu < 2$. The two choices are motivated by physical contents of these states as we shall discuss in greater detail below. Each point in the diagram represents a filling factor $\nu_\mathrm{f}$. Its angular coordinate is $2\pi\nu_\mathrm{f}$ so it falls in the range of (0, 2π) and its radial coordinate is related to its denominator with the outermost point having the smallest value, whereas the origin corresponds to infinite. The innermost points shown here have denominator 51. Even-denominator $\nu$ = 5/2 and 7/2 states are also observed, but they will not be discussed here. Since the energy gap generally decreases as the denominator gets larger, the probability of finding new states diminishes as we move toward the center. The states observed by our measurements are colored red, and other states reported previously in the literature are marked in black. Thanks to the ultrahigh mobility of our sample and the low temperature environment of our demagnetization refrigerator, multiple new FQH states have been discovered. We can compare the results collected in the same fridge without (when the fridge temperature is about 10 mK) and with demagnetization. In the latter case, additional FQH states were observed at 12/23, 13/25, 14/27, 15/29, 16/31, 17/33, 15/31, 14/29, and 13/27 [three of them are indicated in Fig. 1(d)]. The inset in Fig. 1(d) presents a full-field plot, where 4/11 is the last observable FQH state within the magnetic field range accessible in our experiment.

### *Interpretation based on the CF theory*

Since the seminal work of Laughlin[5], theoretical studies of FQH physics have relied heavily on trial wave functions. It is reasonable to begin with the minimal model of electrons in partially filled Landau levels that interact with each other via Coulomb potential. Despite the simplicity of this model, analytical solution is still difficult if not impossible. For a small number of electrons, the many-body Hamiltonian can be solved using exact diagonalization, and the results suggested that liquid instead of solid states are realized at 1/3[56]. However, this does not lead to a comprehensible physical picture. The Laughlin wave function is not only an accurate approximation of the exact ground states, but also played a decisive role in the revelation of fractional charge and fractional braid statistics[6,7]. More generally, the CF theory provides an elegant framework for understanding a large variety of FQH states[11]. Its fundamental postulate is that CFs emerge from a collection of strongly correlated electrons as bound states of bare electrons and an even number of quantized vortices.

To explain the pattern in Fig. 2(a), it is natural to begin with the ones that can be interpreted as IQH states of non-interacting CFs. These states have filling factors $\nu_\mathrm{f} = n/(2pn \pm 1)$ in which $n = 1, 2, 3, \cdots$ that are placed on five colored trajectories corresponding to $2p$ = 2, 4, 6, 8, and 10. The sequence with $2p$ = 2 is very robust and routinely observed in different samples. While the appearance of states at consecutive $n$ is not mandatory, a complete sequence attests to the validity of CF interpretation. As the number of attached fluxes increases, the stability of FQH states deteriorates. For very dilute systems such as $\nu$ = 1/7 and 1/9, FQH states compete with Wigner crystals, and it is not certain which one would prevail in even better samples[57–64]. The maximal filling factor in these sequences is $\nu_{\mathrm{f}}$ = 2/3, so the states at $\nu_{\mathrm{f}}$ > 2/3 should be examined.

For filling factors in the range of 2/3 < $\nu$ < 1, it is tempting to associate them with particle-hole conjugate of the states at $\nu$ < 1/3. However, this is only valid when the electrons have no internal degree of freedom (spin or valley). To account for both one-component and multi-component states, one should still define $\nu = \frac{\tilde{n}}{2\tilde{n}-1}$ with fractional $\tilde{n}$ > 1. For filling factors larger than 1, the results depend on how many internal degrees of freedom are involved. For simplicity, we assume that electrons only carry spin, so $0 \leq \nu \leq 1$ is the spin-down lowest LL and $1 < \nu \leq 2$ is the spin-up lowest LL. If spin-down and spin-up levels are independent, it can be decomposed to an IQH state plus a FQH state at $\nu - 1$. In contrast, when the two spin components are correlated, the state should be mapped to $2 - \nu$ using particle-hole conjugate in the spinful lowest LL. The states at $\nu - 1$ or $2 - \nu$ can be analyzed as before. In some cases, this process may be quite protracted. The 6/5 FQH state is shown as an example in Fig. S1[65].

The states at $\nu_{\mathrm{f}}$ = 4/5, 5/7, 7/9, and 9/11 pose another challenge: they have simple counterparts via particle-hole transformation but they do not belong to the aforementioned CF sequences. We note that the $\nu$ = 9/11 FQH state was observed in a recent experiment[66]. The latter feature is shared by the states at $\nu_{\mathrm{f}}$ = 4/11, 7/11, 4/13, 5/13, 5/17, and 6/17. There were hints about these states in 0 < $\nu$ < 1[67] and the two FQH states at $\nu$ = 4/11 and 5/13 have been firmly established[68,69]. Due to the limited range of magnetic field, our measurement does not reach $\nu$ = 4/13 but reveals a state at $\nu$ = 17/13. It is still possible to explain them using the CF theory. These filling factors can be written as $\nu_{\mathrm{f}} = \tilde{n}/(2p\tilde{n} \pm 1)$ with fractional $\tilde{n}$. For instance, the $\nu_{\mathrm{f}}$ = 4/11 and 5/13 states correspond to $\tilde{n}$ = 4/3 and 5/3. If we assume that the CFs have sufficiently strong interaction, which is generally quite different from the Coulomb interaction between electrons, FQH states of CFs could be realized at the desired $\tilde{n}$ to become FQH states of electrons. This scenario must be examined carefully using numerical calculations because FQH states of CFs could be exotic. For the spin-polarized $\nu_{\mathrm{f}}$ = 4/11 state, it has been proposed that the 1/3 part of the $\tilde{n}$ = 4/3 state is not the Laughlin state[70,71]. Using the mapping to fractionally filled CFs, multi-component states with varying degrees of spin polarization can also be investigated[72]. Both conventional and unconventional ones have been identified at $\nu_{\mathrm{f}}$ = 4/5, 5/7, 4/13, 5/17, and 6/17[73–75]. For a spin-polarized state with simple particle-hole counterpart, the interpretations based on FQH states of CFs and particle-hole conjugate are actually equivalent. It is interesting that the $\nu_{\mathrm{f}}$ < 1/2 regime (from angle 0 to π) has more states than the $\nu_{\mathrm{f}}$ > 1/2 regime. We note that each point in the plot may actually represent multiple states because those at $\nu$ > 1 are mapped to $\nu$ < 1 ones. This distinction between the number of states may be caused by many factors. To reach small filling factors, higher magnetic fields are needed. From an energetic perspective, this enhances the Coulomb energy scale so FQH states are more likely to emerge. However, it is also possible that some filling factors cannot be reached even for the highest accessible magnetic field. In the vicinity of filling factors 0, 1, and 2, nontrivial physics such as the formation of Wigner crystals may overshadow FQH states.

To characterize robustness of the FQH states, we define the ratio $1 - R_{xx}/R_{xx}^{bg}$ to represent the robustness of a FQH state as shown in Fig. 3. The inset in Fig. 3(a) illustrates the same definition of background resistance $R_{xx}^{bg}$ as in Ref.[76]. The extracted values for 0 < $\nu$ < 1 and 1 < $\nu$ < 2 are displayed separately in Fig. 3. For the most prominent CF sequences $n/(2n \pm 1)$ around $\nu$ = 1/2, the ratio exhibits an almost linear dependence on $1/n$ (the $\nu$ = 2/5 and 2/3 states deviate from this trend to some extent). This feature is also observed for the sequences around $\nu$ = 3/4 and $\nu$ = 3/2. It is consistent with the scaling of energy gaps reported in the literature[45,77]. The middle points of these sequences ($\nu$ = 1/2, 3/4, and 3/2) are gapless states in which CFs experience zero magnetic field and form Fermi seas[78]. In contrast, the ratio at $\nu$ = 4/11, 7/11, and 5/13 do not belong to any obvious line, which signifies their special origins. In summary, we have organized the states in Fig. 2 in a systematic way such that their growth pattern becomes transparent. For each colored trajectory, moving from outside to inside traces the fractions in descending order of their robustness.

## *Influence of in-plane magnetic field*

While LLs are generated by an external magnetic field along the perpendicular direction, tilting the field away to create an in-plane component could be useful. In practice, the magnetic field generated by a superconducting solenoid does not change its direction during measurements but samples are rotated. This can be implemented using gear transmission, pressurized liquid $^3$He, and piezo-driven methods[79]. It has been demonstrated that precise control of the tilt angle and excellent reproducibility can be achieved using a piezo-driven rotator in a cryogen-free dilution refrigerator[80], which is also employed in our experiments, as shown in Fig. 1(c). For an ideal 2D system with a fixed perpendicular field, increasing the in-plane magnetic field does not affect orbital motion but increases Zeeman splitting. This has been widely used to induce transitions between FQH states with different spin polarizations[81]. In real quantum wells with finite thickness, the orbital wave functions are also modified and may lead to intricate consequences[82]. The FQH states in the second LL of GaAs exhibit particularly rich features under in-plane magnetic field. In particular, symmetry-breaking stripe and bubble states are often generated[83–85]. It was found by some of us that the energy gap diminishes rapidly at $\nu$ = 14/5, continues to strengthen at $\nu$ = 7/3 and 8/3, and evolves non-monotonically at $\nu$ = 5/2 and 7/2 (increases at first and then decreases)[79,80] when increasing in-plane magnetic field.

Figure 4 presents the longitudinal resistance in tilted magnetic fields at multiple filling factors in the cryogen-free dilution refrigerator. There are several different types of behaviors when the tilt angle $\theta$ and temperature are varied. If the field has no in-plane component, FQH states are observed at $\nu$ = 4/3 and 9/7 but not at $\nu$ = 6/5. As the tilt angle increases to 44.9°, the $\nu$ = 4/3 state diminishes at first but revives after $\theta$ = 41.8°, the $\nu$ = 9/7 state gradually weakens, and the $\nu$ = 6/5 state progressively develops. For a fixed angle at which one state is observed (30.9° or 44.0°), its resistance dip becomes deeper as the temperature is lowered. These results suggest that the $\nu$ = 4/3, 9/7 and 6/5 states are not spin-polarized at $\theta$ = 0° (either partially polarized or fully unpolarized) but the 4/3 state becomes polarized at $\theta$ = 44.9° and the $\nu$ = 6/5 state is spin-polarized at sufficiently large $\theta$. To understand the states at $\nu$ = 4/3 and 9/7, we need to perform a particle-hole transformation in the spinful lowest LLs such that they are mapped to 2 − $\nu$ = 2/3 and 5/7, where spin-unpolarized and partially spin-polarized states of non-interacting CFs exist[65]. For samples with higher carrier density, larger magnetic field is required to reach the $\nu$ = 6/5 state, so it should be observable without an in-plane field. Although the sample temperature in the dilution fridge is considerably higher than that in the nuclear adiabatic demagnetization fridge, the 6/5 state was not found in the latter setting. This fact underscores the merit of an in-plane magnetic field.

Next, we turn to the fragile FQH-like features at $\nu$ = 13/8, 13/11, 18/11, 17/13, and 19/15. The simplest scenario for these states is that the integer parts form inert IQH states in the spin-down lowest LL and the fractional parts form FQH states in the spin-up lowest LL. However, an inspection of the spin and subband degrees of freedom with a large in-plane field reveals other possibilities. The resistance dips at $\nu$ = 19/15 and 17/13 persist up to $\theta$ = 30.9° and disappear at $\theta$ = 44.9°. When the angle further increases, the 17/13 state is no longer observed, but the 19/15 state revives at $\theta$ = 54.9° and disappears again above $\theta$ = 64.9°. This implies that spin transitions occur at $\nu$ = 17/13 and 19/15 when the Zeeman energy is tuned. It contradicts the simplest scenario because in such cases the fractional part is maximally spin polarized. Instead, we should try to explain them by studying the states at 2 − $\nu$ = 9/13 and 11/15, as done before for the $\nu$ = 4/3 state. When the tilt angle is fixed at 44.0° and the temperature is lowered, the resistance dips at $\nu$ = 13/11 and 13/8 become more discernible but the total resistance becomes larger. This is not expected for pure FQH states but points to the existence of insulating background states[60–64]. It is quite likely that FQH states and Wigner crystals or skyrmion crystals compete with each other at these filling factors. While the ground states are insulating crystals, they melt at elevated temperatures such that the system exhibits some characteristics of FQH states. In some cases, these minimum merge into the nearby $\nu$ = 1 or 5/3 plateaus. The even-denominator $\nu$ = 13/8 state is particularly interesting as it requires important revision of the theoretical picture outlined above. If we only consider IQH states or odd-denominator FQH states of CFs, the FQH states of electrons necessarily have odd denominators. It has been proposed that FQH states could arise at $\nu$ = 3/8 and 3/10[73,86,87] when chiral $p$-wave pairing of CFs generates Moore-Read-type states[51]. While some signatures have been reported in GaAs electron systems, their presence has yet to be firmly confirmed[67,88]. For example, potential CF fermi liquid may also appear as a minimum in the longitudinal resistance. In contrast, convincing observations have been reported in GaAs hole systems, and LL mixing is conjectured to play an important role[35]. The 13/8 dip in our sample occurs at a rather low magnetic field so LL mixing is also strong. It should be emphasized that the details about LL mixing are quite different in electron and hole systems. There is strong spin-orbit coupling in GaAs hole systems, so that neither LL index nor the spin is a good quantum number, and the spatial component of the single-particle wavefunction consists of a mixture of different Landau oscillators[89]. For example, two levels may be brought very close to each other so they have substantial mixing with each other, whereas other LLs are more distant in energy. In GaAs electron systems, LLs with the same spin but different orbital indices are equally spaced.

It is evident that applying an in-plane magnetic field allows one to further probe the nature of putative FQH states and may even generate additional ones that are absent without the in-plane field. It has been reported in double-layer systems that FQH states may spontaneously break the layer symmetry and form imbalance states such as 11/15 = 2/5 + 1/3, 9/13 = 3/13 + 6/13, etc. It is likely that a four-component state with spin and subband/layer symmetry may give rise to this complex evolution[90,91]. To provide further insights into the measured data at $\nu$ = 13/11, 18/11, 17/13, and 19/15, we check if they could be detected by numerical calculations. This is a very demanding task so our results are rather preliminary. Even if these states do exist, their energy gaps would be quite small. This means that minor changes in the Hamiltonian might destabilize them, so it is imperative to construct an accurate model for such a system. Two factors that have important quantitative effects on the energy gaps are sample thickness and LL mixing. The former is treated using an infinite square well potential and the latter is tackled by the method proposed in Ref.[92]. We have only studied the states belonging to the simplest scenario mentioned above using exact diagonalization. For two-component states that involve both spins, there are severe limitations due to the rapid growth of the Hilbert space dimension. The only active part is the partially filled spin-up lowest LL with renormalized Coulomb interaction. As indicated by softening of the collective modes, the system is likely compressible at $\nu$ = 13/11, 18/11, and 17/13. This is consistent with the measurements and reinforces the conjecture that the 17/13 state is two-component. The calculation at $\nu$ = 19/15 is rather challenging and we lack definitive results, so the property of this state is deduced primarily based on the experimental fact that it does not survive when the in-plane magnetic field gets sufficiently large. For more details about numerical calculations, see Section 3 of the Supplemental Material.

### *FQH states in other platforms*

In addition to electron-doped GaAs quantum wells, FQH states have been observed in many other platforms. It is challenging to provide a comprehensive summary of the existing data here, so we select four materials that have certain noteworthy features: hole-doped GaAs[33,34], monolayer graphene[23–26], BLG[28,29,32], and monolayer $WSe_2$[31]. The energy bands and single-particle Landau orbitals are quite different, LL mixing could be strong, and bare Coulomb potential should be properly modified to account for screening etc. Despite these differences, the majority of FQH states in these systems still have odd denominators, as summarized in Figs. 2(b)-2(e). We also present the observed fractions in Table 1. The CF theory still provides a good starting point to understand these results. However, special care must be taken in some cases where other candidates are also plausible. For example, non-Abelian Read-Rezayi states may be realized at $\nu_f$ = 2/5 and 3/5[93]. FQH states of CFs are scarcer, which may be due to insufficient sample qualities but also calls for more theoretical investigations.

When the observed fractions deviate from the pattern discussed above, it often implies interesting results regarding the nature of some FQH states. An interesting subclass is even-denominator ones that probably support non-Abelian anyons. For example, a few half-filled states have been observed in BLG[28,29,32]. In contrast to the linear Dirac dispersion of monolayer graphene, BLG has quadratic bands. The single-particle wave functions in the LLs are generally complicated superpositions of non-relativistic Landau orbitals. By tuning external magnetic and vertical electric fields, the LLs exhibit sophisticated behaviors. For a wide range of parameters, the N = 0 and 1 LLs are approximately degenerate that span the filling factor range –4 to 4. Zibrov *et al.* observed an FQH state at $\nu$ = 3/2 with rather large energy gap and induced phase transitions by tuning the electric field[28]. Li *et al.* reported states at $\nu$ = –5/2, –1/2, 3/2, and 7/2 in dual-gated devices, and studied the evolution of energy gaps with external fields[29]. Furthermore, by applying an in-plane magnetic field, the states were found to be spin-polarized. Huang *et al.* found another state $\nu$ = 5/2 and discussed its spontaneous valley polarization[32]. It is believed that all half-filled states occur in the N = 1 LL of BLG. Its single-particle orbitals are similar to those in the second LL of electron doped GaAs, so the FQH states are very likely of the Moore-Read type[52,53]. Depending on the details of LL mixing, they could be Pfaffian, anti-Pfaffian, or particle-hole symmetric Pfaffian states. Their presence also affects the patterns of odd-denominator states. In some cases, FQH states were observed at either $\nu_f$ = 6/13, 9/17 or 7/13, 8/17 but not all of them[32]. This could be explained as the formation of Levin-Halperin daughter states associated with Pfaffian or anti-Pfaffian states instead of CF states[94]. Another indication of their non-CF origin is that no FQH states were observed at lower denominator fractions 6/11 and 7/15. The rich phenomenology, excellent tunability, and large energy gaps of BLG made it a promising platform to further advance the investigations of non-Abelian anyons. For TMDs, the band structures are described by massive Dirac fermion with spin valley locking. The N = 1 LL is very similar to that in GaAs, but the single-particle orbitals have a small weight in the non-relativistic zeroth LL. Shi *et al.* reported a $\nu$ = 3/2 FQH state in monolayer $WSe_2$, which is likely of the Moore-Read type[31]. One interesting feature is that the LL mixing parameter (interaction energy divided by the cyclotron gap) is quite large, so there may be nontrivial consequences.

Recent years have also witnessed rapid progress on FQH physics without magnetic field, namely fractional quantum anomalous Hall (FQAH) states or fractional Chern insulators (FCIs)[39–43]. In twisted bilayer $MoTe_2$, FQAH states at $\nu$ = –2/3 and –3/5 were observed using optical, thermodynamic, and transport methods[39–42]. In rhombohedral pentalayer graphene, FQAH states at $\nu$ = 2/3, 3/5, 4/7, 4/9, 3/7 and 2/5 were observed in transport measurements[43]. It is remarkable that the filling factors in both cases still belong to the simplest Jain sequence. Indeed, numerical calculations found that these FQAH states should have the same universal topological properties as their counterparts in LLs[95]. Nevertheless, the microscopic energetics may be quite intricate and the elementary excitations may acquire very different characteristics.

## Discussion and conclusion

In summary, we have proposed a pattern for the FQH states in ultrahigh-quality GaAs electron systems. The filling factors of FQH states are organized into a visual pattern and comparisons of several different materials are made. This serves as an overview of the state of the art that could be very useful for future reference. Based on the CF theory, the nature of these states is analyzed. A few of them seem to be quite exotic and warrant more in-depth studies. We have also attempted to explain the data using the hierarchy theory[96,97] but the outcome is less satisfactory as shown in Fig. S2[20,49,50,57,62,64,66–69,71,96,98–107]. Looking into the future, we expect that FQH physics will thrive for many more years. To probe FQH states with very small energy gaps, a lower temperature environment should be pursued. However, the development of ultra-low-temperature refrigeration since the beginning of the 21st century has mostly focused on adapting to cryogen-free pre-cooling technology. When it comes to reducing the lowest attainable temperature, we must frankly admit that physicists are still waiting for a fundamental advance to go beyond nuclear adiabatic demagnetization. Besides electrical transport properties, other aspects of FQH states could also be interesting. Some experimental methods have been developed for this purpose and it is desirable to invent new ones. The integration of experimental probes with a better low-temperature environment would also be very helpful. Finally, the significance of making better-quality samples and finding new platforms can never be overemphasized. As the history of low-temperature physics has shown, many breakthroughs were completely unanticipated, suggesting there must be other exciting phenomena not within our imagination.

## Methods

***Sample information and measurement conditions.*** We perform electrical transport measurements on a GaAs/AlGaAs heterostructure. The GaAs well has a width of 49 nm, so electrons are confined to the lowest subband and the second subband is not populated[20]. The sample is prepared in a Van der Pauw structure with size 2.5 mm × 2.2 mm and ohmic contacts are made from annealed InSn alloy. Our measurements are carried out in a nuclear adiabatic demagnetization refrigerator with base temperature below 1 mK and other fridges as shown in Figs. 1(a)-1(c). Low-temperature RC filters and silver-epoxy filters are employed to minimize the electron temperature. Based on the experimental data, we conclude that electron temperatures below 12 mK were reached. By comparing our experimental setup with that in Ref.[107], it is plausible that the electron temperature is higher than 4 mK. More details about the filtering scheme are given in Section 4 of the Supplemental Material. It is found that the mobility of our sample is $3.7\times10^7\ \mathrm{cm^2V^{-1}s^{-1}}$, which is comparable with the recently published value $4.4\times10^7\ \mathrm{cm^2V^{-1}s^{-1}}$ [20].

The base temperature of the fridge with rotating stage shown in Fig. 1(c) is 11 mK, while the measurement apparatus is installed on a probe such that both the sample holder and the electrons are at much higher temperatures. In addition, another rotatable mechanical sample holder has been constructed using only oxygen-free copper adapters whose angle can be tuned in the range of 0 to 90 degrees and allows us to access electron temperature lower than 40 mK.

***Data analysis.*** To estimate the areal density of electrons, the most robust FQH states at $\nu$ = 2/5, 3/7, 4/7, 7/5, 10/7, 11/7, and 8/5 are considered. By fitting them with the associated magnetic field values, we conclude that the density is about $1.0\times10^{11}$ $\text{cm}^{-2}$ with an uncertainty less than $3.9\times10^{7}$ $\text{cm}^{-2}$. Using the fitted density value and the magnetic field at which $R_{xx}$ reaches the local minimum, each filling factor can be computed numerically and then compared with the theoretical value. This leads to a relative uncertainty less than ±0.4% in the determination of filling factors. It is still possible to definitely separate two close fractions given this upper limit. For fraction 15/31, within a range of $(1\pm0.4\%)\times15/31$ and with denominators limited to less than 51, the possible fractions are 14/29, 15/31, 16/33, and 17/35. Among these, 15/31 (0.4839) is the closest approximation to the filling factor of 0.4838 derived from the magnetic field of minimum $R_{xx}$ in the $R_{xx}-B$ curve based on the fitted density. The exact diagonalization calculation was applied to the in-plane magnetic field data, where the lowest LL with spin-down and the second LLs with both spins are integrated out using screening theory and all higher LLs are integrated out using perturbative calculations[92].

## Acknowledgements

We thank Bo Yang, Xin Wan, Hailong Fu, Yuzhu Wang, Xincheng Xie, Chenjie Wang, Jiasen Niu and Kun Yang for discussions.

## Funding

This work was supported by the National Key Research and Development Program of China (2021YFA1401900) and the National Natural Science Foundation of China (12141001). Ying-Hai Wu was supported by the National Natural Science Foundation of China (12174130), Wenchen Luo was supported by the Hunan Provincial Natural Science Foundation of China (2025JJ50019), and Wei Zhu was supported by the National Natural Science Foundation of China (12474144). This work was supported by Quantum Science and Technology–National Science and Technology Major Project (2021ZD0302600). Loren Pfeiffer was supported by the Gordon and Betty Moore Foundation's EPiQS Initiative (GBMF9615.01).

## Author contributions

S.Y. and X.L. conceived the project. Z.Chen, Y.Z., Z.Cui, S.Y., and X.L. performed the measurements. J.Y. and X.L. built the nuclear adiabatic demagnetization refrigerator. L.N.P., K.W.W., K.W.B., and A.G. contributed high-quality samples. Z.Chen, Y.L., W.Z., W.L., Y.-H.W., S.Y., and X.L. analyzed the data. Z.Chen, Y.-H.W., S.Y., and X.L. wrote the paper with suggestions from the others.

**Conflict of interest statement.** None declared.

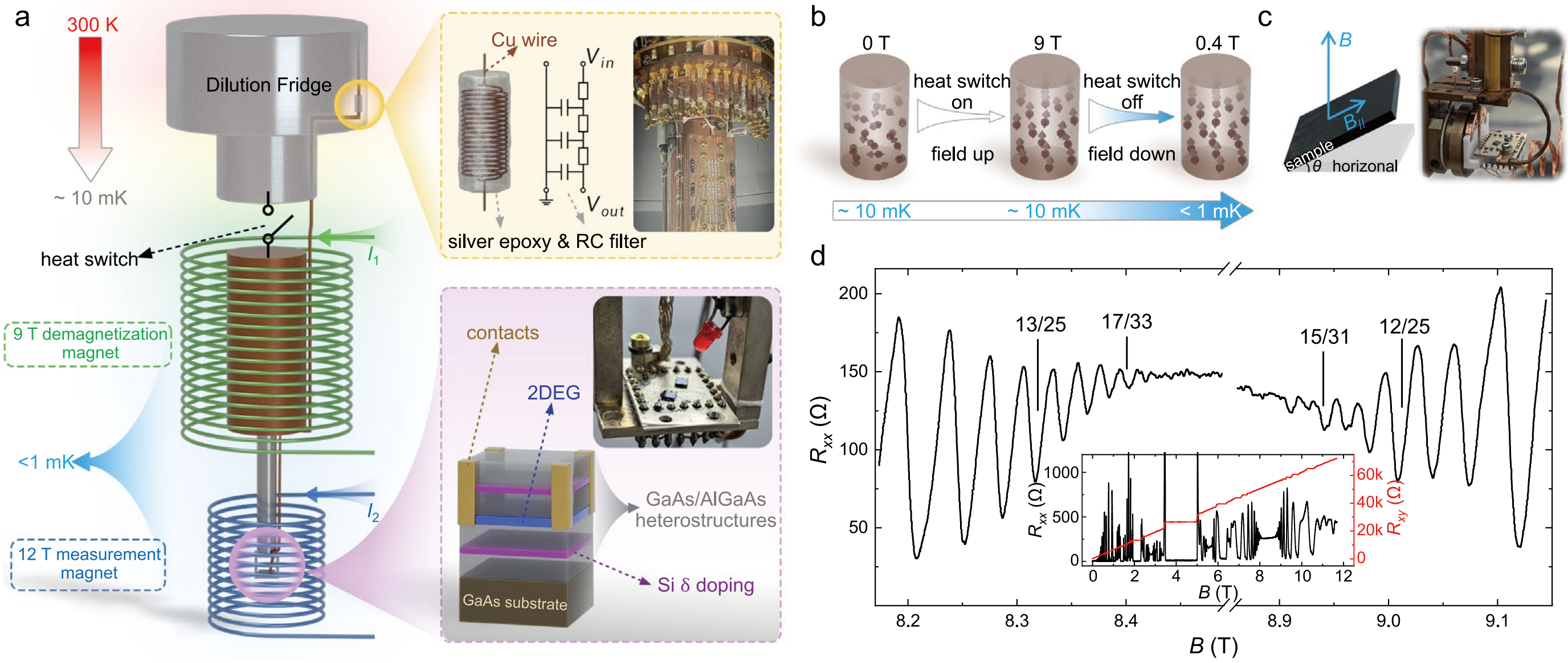

**Figure 1.** Measurement apparatus and magnetoresistances. (a) Configuration of the nuclear adiabatic demagnetization fridge in which the data in (d) are taken at an electron temperature estimated to be less than 12 mK. Electrical filters and structure of the GaAs/AlGaAs quantum well are displayed in the zoom-in boxes. (b) Schematic of the adiabatic demagnetization process of copper nuclear spins indicated by direction of arrows. (c) Schematic and photograph of the sample rotator in a cryogen-free dilution fridge. (d) FQH states in a sample with electron density $\rho = 1.0\times10^{11}$ cm$^{-2}$ and mobility $\mu = 3.7\times10^{7}$ cm$^{2}$V$^{-1}$s$^{-1}$. Minima of the longitudinal resistance $R_{xx}$ are observed at many filling factors up to $\nu$ = 17/33 and $\nu$ = 15/31. Traces of $R_{xx}$ and the Hall resistance $R_{xy}$ in the magnetic field range 0 ~ 11.7 T are shown in the inset.

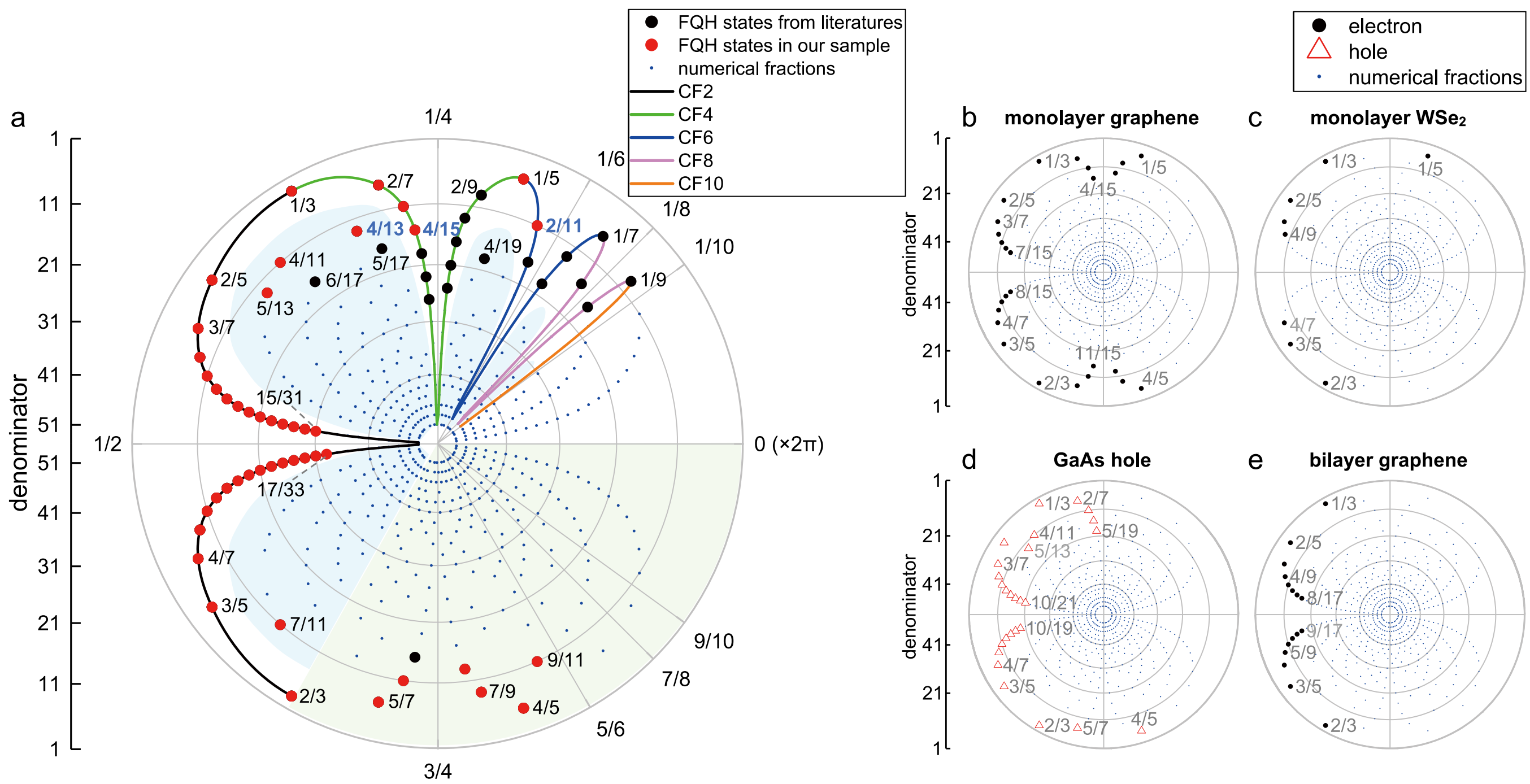

**Figure 2.** Circular pattern of FQH states. Each filling factor $\nu = p/q$ is represented by a dot in a polar form with angle $\theta = 2\pi p/q$. The radial axis represents the denominator q in an inverted manner: q = 1 is at the outermost edge, and larger denominators lie progressively closer to the center (up to q = 55). In general, the distance to the origin is chosen as $r(q) = 55 - q$. As a consequence, simpler fractions appear near the rim, whereas higher-denominator fractions cluster toward the center. For example, $\nu = 1/3$ is located at angle $2\pi/3$ and radius 52. (a) Red dots are FQH states found in our measurements, black dots are FQH states that have been reported in the literature[20,49,50,57,62,64,66–69,98–107], and blue dots are all possible fractions with odd denominators smaller than 52 but not yet observed in experiments. Blue text is used to indicate three newly observed FQH states at $\nu = 17/13$, 19/15, and 13/11. Black, green, blue, pink, and orange lines are used to connect fractions that correspond to IQH states of CFs with 2, 4, 6, 8, and 10 attached fluxes, respectively. Blue shaded areas enclose the fractions that cannot be understood as IQH states of CFs. Green shaded area encloses $\nu > 2/3$ fractions that may be particle-hole conjugates of certain states at $\nu < 1/3$ or FQH states of CFs. (b-e) FQH states in four other systems[33,55,108,109] are organized using the same pattern as in panel (a). Actual filling factor range is $0 < \nu < 2$ in (c) and $0 < \nu < 1$ in panel (b, d, e).

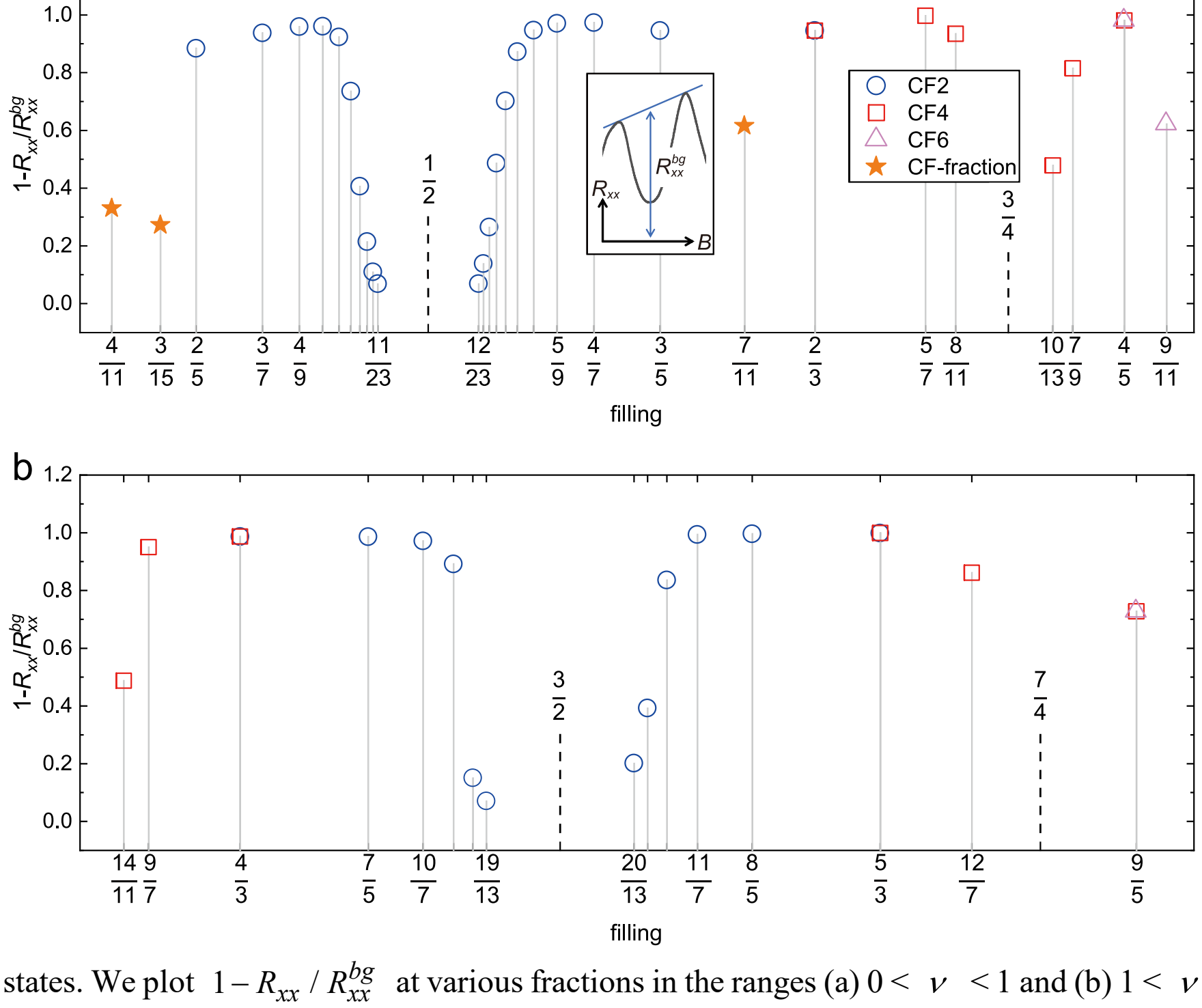

**Figure 3.** Strength of FQH states. We plot $1-R_{xx}/R_{xx}^{bg}$ at various fractions in the ranges (a) $0<\nu<1$ and (b) $1<\nu<2$. $R_{xx}^{bg}$ of a FQH state is the background resistance as defined by the inset of panel (a). Hollow blue circles, red squares, and pink triangles are FQH states that correspond to IQH states of CFs with 2, 4, 6 attached fluxes, respectively. Solid stars are FQH states that cannot be associated with IQH states of CFs. The data is collected in one single cool down of the nuclear adiabatic demagnetization fridge.

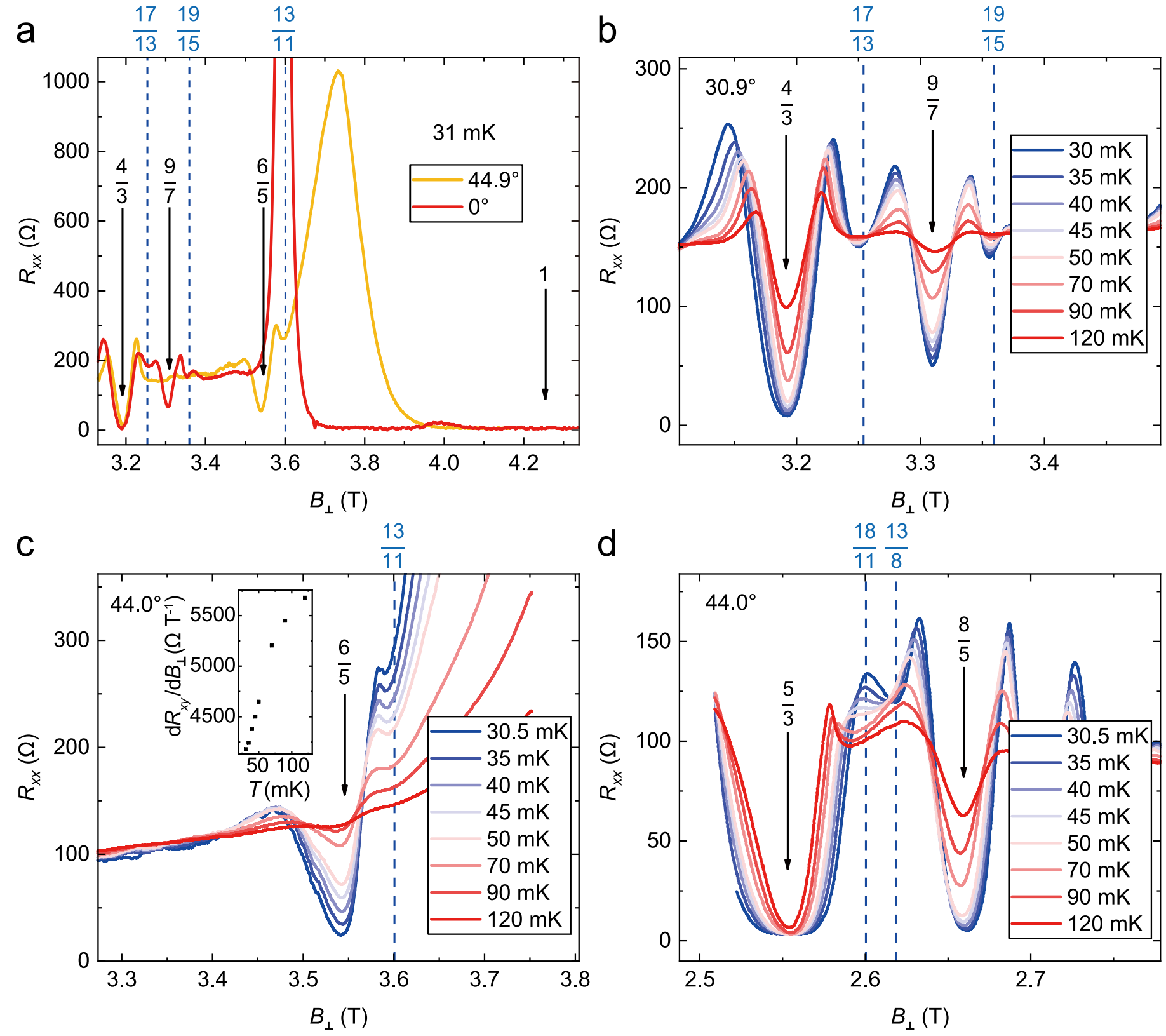

**Figure 4.** Evolution of FQH states with in-plane magnetic field. $R_{xx}$ is plotted versus the perpendicular magnetic field $B_\perp$ at different tilt angle $\theta$ and temperature $T$. In each panel, a few interesting fractions are indicated. (a) $\theta = 0°$ and 44.9° at $T = 31$ mK. FQH states at $\nu = 4/3$, 6/5, and 9/7 exhibit different behaviors when $\theta$ is tuned. (b) $\theta = 30.9°$ with $T$ tuned from 30 mK to120 mK. FQH states appear at $\nu = 17/13$ and 19/15. (c) $\theta = 44.0°$ with $T$ tuned from 30.5 mK to 120 mK. The resistance dip at $\nu = 13/11$ becomes more pronounced at lower temperature but the overall value also increases. Its inset shows the derivative $dR_{xy} / dB_\perp$ of the Hall resistance at $\nu = 13/11$ and $\theta = 44.0°$. This provides support for it being a FQH state. (d) $\theta = 44.0°$ with $T$ tuned from 30.5 mK to 120 mK. The resistance dips around $\nu = 13/8$ and 18/11 are similar to the one at $\nu = 13/11$ but their locations shift with temperature. All measurements in tilted fields were performed using a dilution fridge with a minimum electron temperature of approximately 40 mK.

**Table 1.** List of odd-denominator FQH states that have been reported[20,49,50,57,62,64,66–69,98–107]. The filling factors are divided into multiple groups according to the trajectories in Fig. 2. For numbers that are larger than 2/3, we convert them to $\nu = N \pm n/(2pn \pm 1)$ with $N, p, n =$ 1, 2, 3, ⋯.

| FQH states | formula | $n$ |
|---|---|---|
| 1/3, 2/5, 3/7, 4/9, 5/11, 6/13, 7/15, 8/17, 9/19, 10/21, 11/23, 12/25, 13/27, 14/29, 15/31 | $\nu = n/(2n+1)$ | {1, ⋯, 15} |
| 2/3, 3/5, 4/7, 5/9, 6/11, 7/13, 8/15, 9/17, 10/19, 11/21, 12/23, 13/25, 14/27, 15/29, 16/31, 17/33 | $\nu = n/(2n-1)$ | {2, ⋯, 17} |
| 1/5, 2/9, 3/13, 4/17, 5/21, 6/25 | $\nu = n/(4n+1)$ | {1, ⋯, 6} |
| 2/7, 3/11, 4/15, 5/19, 6/23, 7/27 | $\nu = n/(4n-1)$ | {2, ⋯, 7} |
| 1/7, 2/13, 3/19 | $\nu = n/(6n+1)$ | {1, 2, 3} |
| 2/11, 3/17 | $\nu = n/(6n-1)$ | {2, 3} |
| 1/9, 2/17 | $\nu = n/(8n+1)$ | {1, 2} |
| 2/15 | $\nu = n/(8n-1)$ | {2} |
| 5/7, 8/11, 11/15 | $\nu = 1 - n/(4n-1)$ | {2, 3, 4} |
| 4/5, 7/9, 10/13 | $\nu = 1 - n/(4n+1)$ | {1, 2, 3} |
| 9/11 | $\nu = 1 - n/(6n-1)$ | {2} |
| 13/11 | $\nu = 1 + n/(6n-1)$ | {2} |
| 6/5, 11/9 | $\nu = 1 + n/(4n+1)$ | {1, 2} |
| 9/7, 14/11, 19/15 | $\nu = 1 + n/(4n-1)$ | {2, 3, 4} |
| 4/3, 7/5, 10/7, 13/9, 16/11, 19/13, 22/15 | $\nu = 1 + n/(2n+1)$ | {1, ⋯, 7} |
| 5/3, 8/5, 11/7, 14/9, 17/11, 20/13, 23/15 | $\nu = 1 + n/(2n-1)$ | {2, ⋯, 8} |
| 12/7 | $\nu = 2 - n/(4n-1)$ | {2} |
| 9/5 | $\nu = 2 - n/(4n+1)$ | {1} |
| 11/5 | $\nu = 2 + n/(4n+1)$ | {1} |
| 16/7 | $\nu = 2 + n/(4n-1)$ | {2} |
| 7/3, 12/5, 22/9, 32/13 | $\nu = 2 + n/(2n+1)$ | {1, 2, 4, 6} |
| 8/3, 13/5, 23/9 | $\nu = 2 + n/(2n-1)$ | {2, 3, 5} |
| 19/7 | $\nu = 3 - n/(4n-1)$ | {2} |
| 14/5, 25/9 | $\nu = 3 - n/(4n+1)$ | {1, 2} |
| 16/5 | $\nu = 3 + n/(4n+1)$ | {1} |
| 10/3 | $\nu = 3 + n/(2n+1)$ | {1} |
| 11/3 | $\nu = 3 + n/(2n-1)$ | {2} |
| 19/5 | $\nu = 4 - n/(4n+1)$ | {1} |
| 21/5 | $\nu = 4 + n/(4n+1)$ | {1} |
| 24/5 | $\nu = 5 - n/(4n+1)$ | {1} |
| 4/19, 5/17, 4/13, 6/17, 4/11, 5/13, 7/11, 17/13 | These fractions cannot be written as $\lvert \nu - N \rvert = n/(2pn \pm 1)$ using integer $N$. | |

# Supplemental Material for

## Cascade of fractional quantum Hall states in 2D system

Zhimou Chen, Jiaojie Yan, Yuxuan Zhu, Zhe Cui, Loren N. Pfeiffer, Kenneth W. West, Kirk W. Baldwin, Adbhut Gupta, Yang Liu, Wei Zhu, Wenchen Luo, Ying-Hai Wu*, Shuai Yuan* and Xi Lin*

*Correspondence should be addressed to Y.-H.W. (yinghaiwu88@hust.edu.cn), S.Y. (shuaiy81@uw.edu) and X.L. (xilin@pku.edu.cn)

### 1. Illustrations of two possible $\nu = 6/5$ states

The 6/5 FQH state could be an IQH state in the spin-down lowest LL plus a 1/5 FQH state in the spin-up lowest LL. The latter is the $\tilde{n} = 1$ IQH state of CFs with four fluxes attached as shown in Fig. S1. If it is a two-component state, we first apply a particle-hole transformation to obtain a state at $2 - \nu = 4/5$. It is mapped to a $\tilde{n} = -4/3$ FQH state of CFs with two fluxes attached. Another particle-hole transformation relates 4/3 to the spin-singlet Jain state at 2/3 [1].

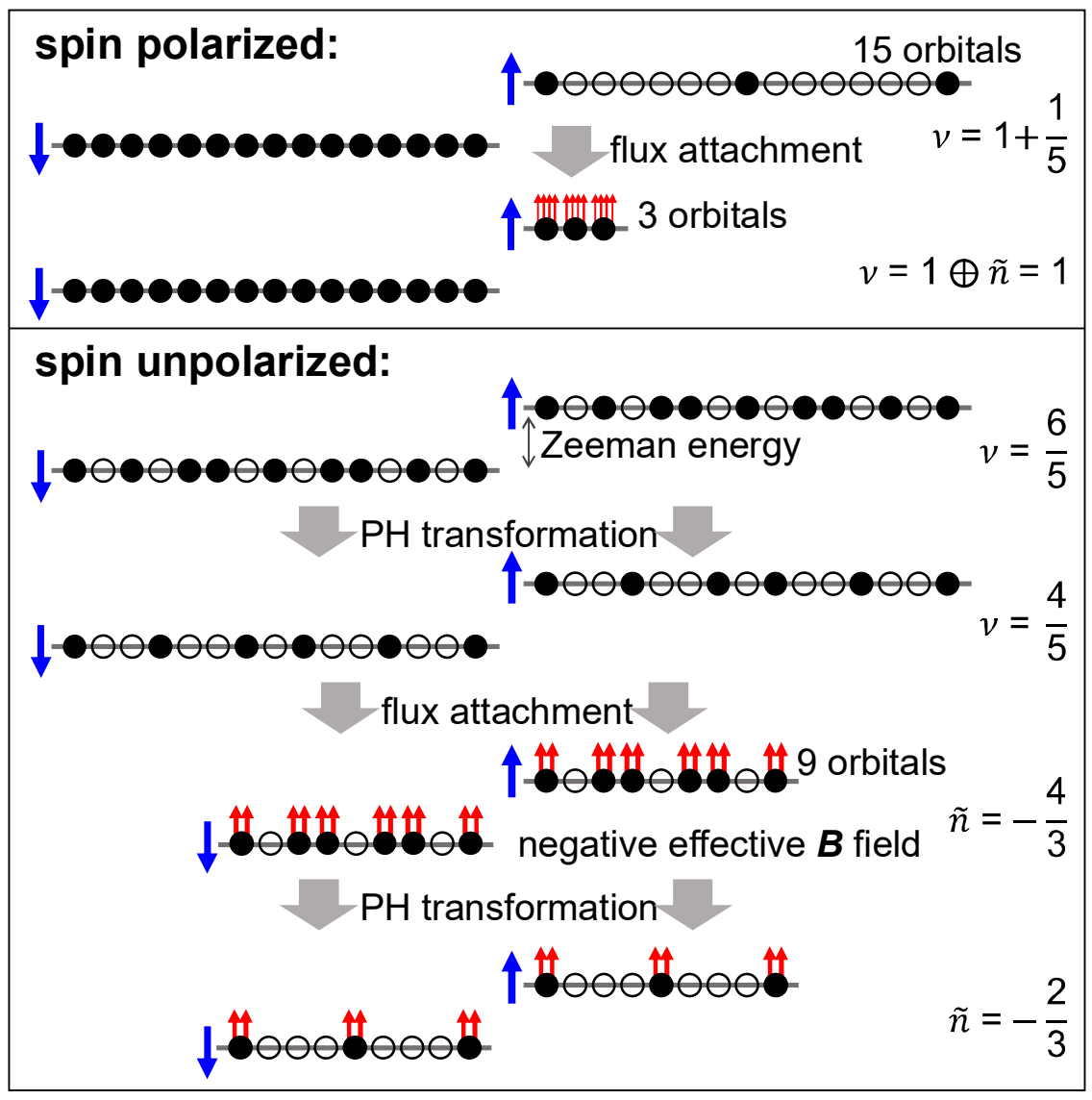

**Figure S1.** Schematics of two possible $\nu = 6/5$ FQH states. In both cases, each LL has 15 orbitals and the number of electrons is 18. Blue arrows represent the spin direction of LLs, solid circles represent electrons, hollow circles represent vacant orbitals, and red arrows represent flux attached to electrons. Details about the constructions are given in the text.

### 2. Interpretation of experimental data using the hierarchy theory

The basic assumption of hierarchy theory is that the elementary charged excitations of one IQH or FQH state may form an incompressible state such that the whole system enters another FQH state[2]. Its starting point is usually taken as the $\nu = 1$ IQH state or $\nu = 1/m$ Laughlin state. For example, a $\nu = 2/5$ state is obtained when the quasiparticles of the 1/3 state form a bosonic $\nu = 1/2$ Laughlin state. For spin-polarized electrons, this approach predicts that FQH states may be found at

$$\nu = \cfrac{1}{m + \cfrac{\alpha_1}{p_1 + \cfrac{\alpha_2}{p_2 + \dots}}} \tag{1}$$

with $m = 1, 3, 5, \cdots$ , $\alpha_n = 0, \pm 1$, and $p_n = 2, 4, 6, \cdots$ . The parameters in this formula are collected as $[m, \alpha_1, p_1, \cdots, \alpha_n, p_n]$.

We provide a sketch about how to explain the data in Fig. S2 using the hierarchy theory. For simplicity, only spin-polarized states are considered here. One principal sequence begins at $\nu = 1/3$ and traverses those at 2/5, 3/7, $\cdots$, $n/(2n+1)$. Another principal sequence begins at $\nu = 1$ and traverse those at 2/3, 3/5, $\cdots$, $n/(2n-1)$. While they terminate at $\nu = 1/2$, the hierarchy theory does not reveal the presence of composite fermion liquid. It is more interesting to study the states at $\nu = 4/11$, 4/13, and 5/13. For example, 4/11 can be obtained if the quasiparticles of the 1/3 state form a bosonic $\nu = 1/4$ Laughlin state. Its topological properties are expected to be the same as the FQH state of CFs in which the fractional part form a Laughlin state. However, previous works have found that the CFs may form an unconventional 1/3 state[3], and it is not obvious how to construct its equivalent in the hierarchy framework.

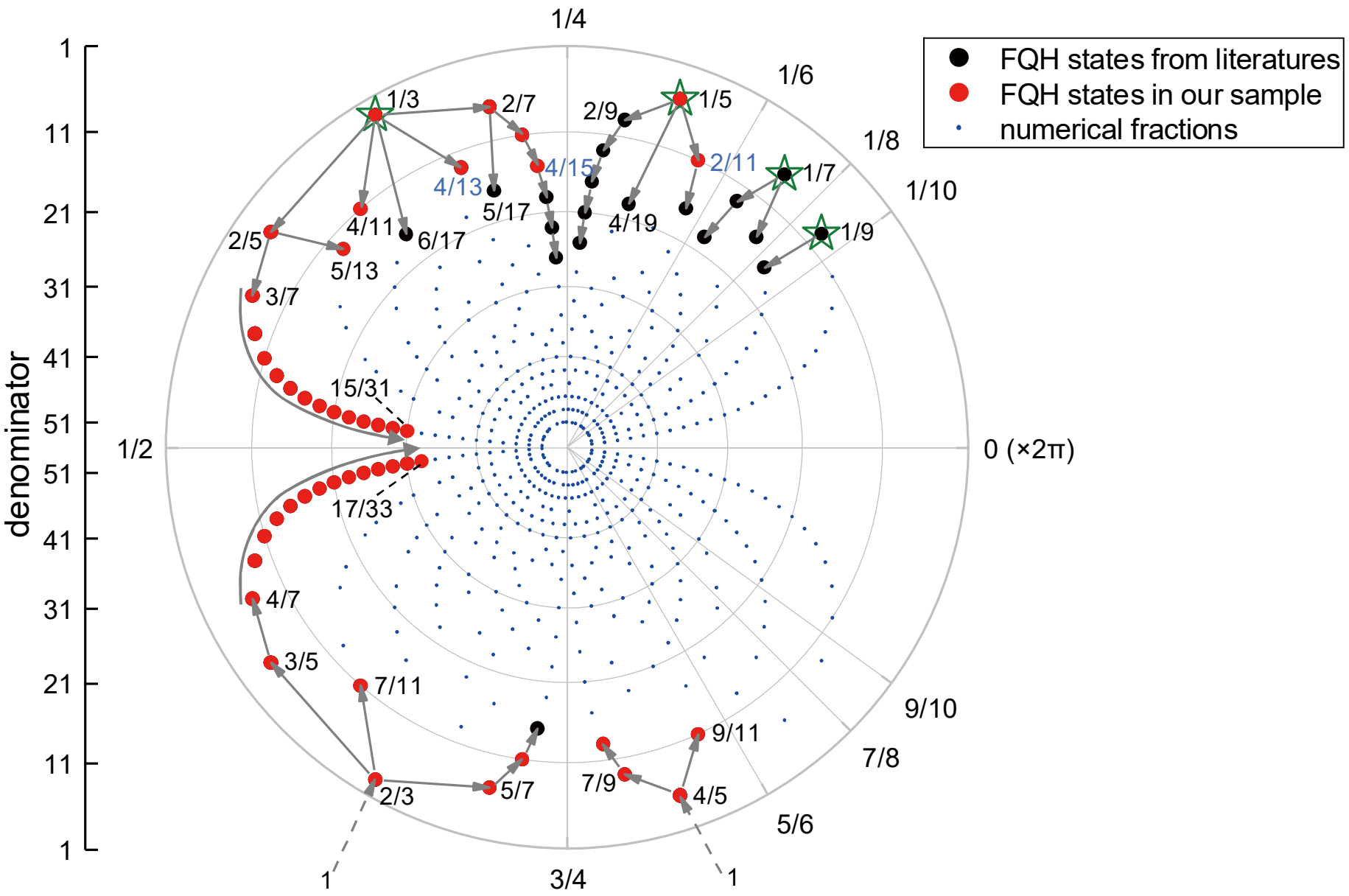

**Figure S2.** Interpretation of experimental data using the hierarchy theory. The same data points as in Fig. 2(a) are plotted[4–23]. Hollow green stars indicate the starting points of several hierarchies. Gray arrow lines indicate the path from a parent $[m, \alpha_1, p_1, \cdots, \alpha_{n-1}, p_{n-1}]$ to its daughter $[m, \alpha_1, p_1, \cdots, \alpha_{n-1}, p_{n-1}, \alpha_n, p_n]$. For the $\nu = 2/3$ and 4/5 states, $\nu = 1$ IQH state is their starting point and connected to them by dashed gray arrow line.

### 3. Numerical results and energy gaps at several filling factors

Exact diagonalization (ED) is a standard tool for studying FQH physics. It has played a decisive role in elucidating the nature of FQH states since the ground-breaking work of Laughlin. The basic idea of ED is straightforward. When a finite number of electrons are placed in a finite number of Landau orbitals, we construct all the Fock states corresponding to non-interacting Slater determinants. The many-body Hamiltonian is written in second quantized form and converted to a numeral matrix in the Fock states basis. If the system has symmetries, this matrix can be organized into a block diagonal form such that each block is treated separately. Using iterative sparse matrix eigensovler such as the Lanczos method, a few low-lying eigenvalues and eigenstates of the Hamiltonian matrix can be obtained. One then proceeds to examine energy gaps, wave function overlaps, and entanglement properties etc. In general, ED can only be done for a small number of electrons because the Hilbert space dimension grows exponentially.

We have studied the $\nu$ = 13/11, 17/13, and 19/15 states using ED. The integer parts form IQH states whereas the fractional parts are assumed to be spin-polarized. It is much more difficult to include the spin degree of freedom because this would greatly enlarge the Hilbert space. To account for LL mixing effects, we employ both random phase approximation (RPA) and perturbation theory. The dielectric function in RPA is obtained by integrating out the neighboring LLs. In perturbative treatment of the contributions from higher LLs, two-body interactions are renormalized by zero-sound as well as Bardeen-Copper-Schrieffer diagrams, and three-body interactions arising from two screened vertices are considered[24]. The finite thickness of quantum wells is modeled by an infinite square well potential. Aided by these approximations, we obtain an effective single-LL description of the problem with a renormalized Coulomb interaction.

As shown in Fig. S3, the ground states for 8 electrons at $\nu$ = 13/11, 17/13 locate at $q \neq 0$ so the systems are compressible. This is consistent with the experimentally observed disappearance of the $R_{xx}$ dip at high in-plane magnetic fields. For $\nu$ = 19/15, preliminary numerical results suggest that an incompressible state may be stabilized. However, this claim should be viewed with caution because the system size is severely limited: only four momentum sectors are available. The true nature of the system can only be ascertained by more experimental investigations. At high in-plane magnetic fields, the $R_{xx}$ dip at $\nu$ = 19/15 disappears. This may be caused by a transition into a compressible state or because the energy gap is too small to be detected within our measurement capabilities.

We have performed Arrhenius fittings of the experimental data at different temperatures using the same method as in Ref.[25]. For tilt angle 30.9°, our fittings yielded energy gaps of approximately 32 mK at $\nu$ = 19/15 and 26 mK at $\nu$ = 17/13. These numbers are comparable to the base electron temperature of our rotatable sample stage (approximately 40 mK). This proximity in energy scales explains why we only observed weak dips in the $R_{xx}$ traces of Fig. 4(b). As for the $\nu$ = 13/11 state, the exceptionally large insulating background resistance disallows a valid Arrhenius fitting.

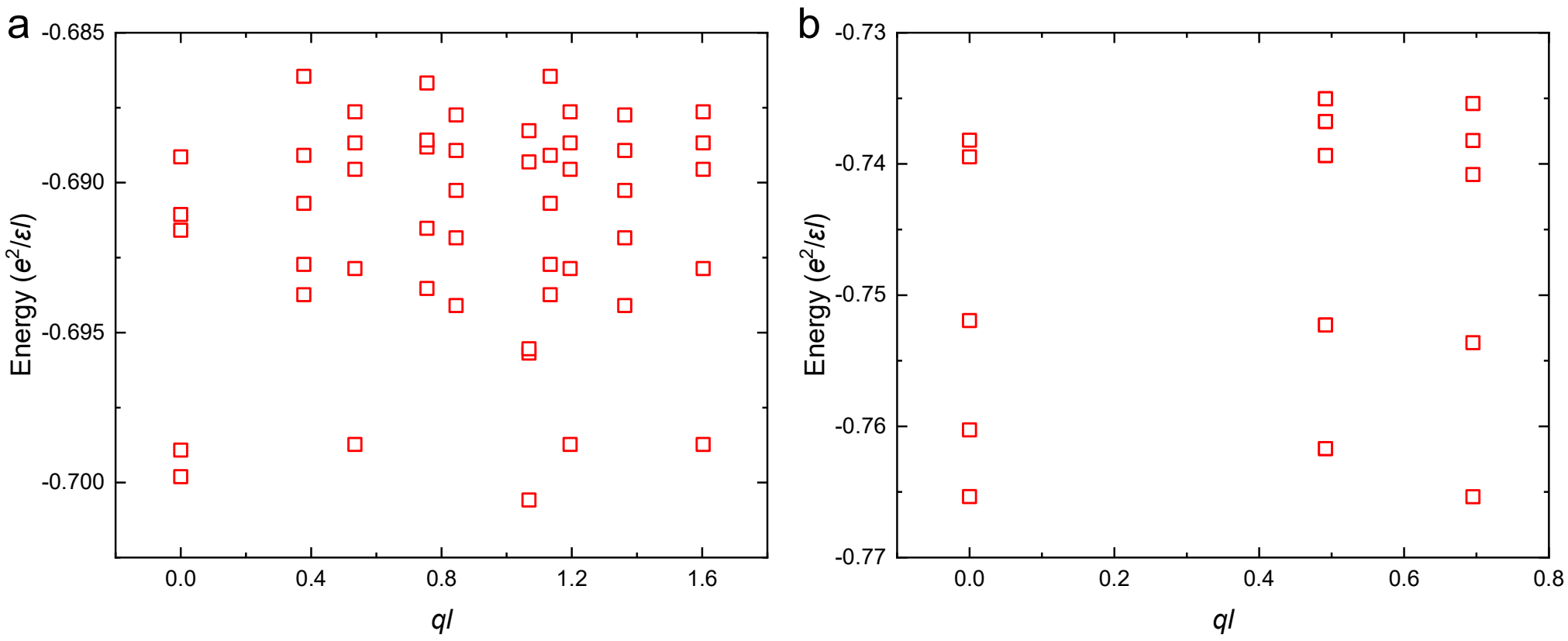

**Figure S3.** Numerical results for the (a) $\nu$ = 13/11 and (b) $\nu$ = 17/13 states with 8 spin-polarized electrons.

### 4. Filtering and electron temperature

The filters in the measurement leads play a crucial role in lowering the electron temperature of the sample. Our nuclear adiabatic demagnetization refrigerator is primarily used for quasi-DC measurements. Consequently, we have installed low-pass filters in the signal lines. We incorporated home-made silver-epoxy filters and RC filters in each measurement line. These two types of filters are connected in series with each other.

The filtering performance of our silver-epoxy filter and RC filter is consistent with the results reported in Ref.[26]. The attenuation is essentially better than -60 dB above 1 MHz and exceeds -90 dB above 1 GHz. Our silver epoxy filter is composed of a 0.1 mm diameter copper enameled wire encased in silver metal powder, and it operates primarily by utilizing the skin effect to filter high-frequency electromagnetic waves. The RC filter is a third-order configuration with respective resistor and capacitor values for each stage of (510 Ω, 4.7 nF), (820 Ω, 2.2 nF), and (1500 Ω, 1.1 nF). The resistors and capacitors are mounted on an insulating substrate made of sapphire to ensure reasonable thermal contact with the cold plate. Due to the high resistor values, which could generate significant thermal noise at room temperature, the RC filter is mounted on the mixing chamber plate.

Based on the performance of the same sample in different refrigerators, we estimate that the electron temperature of the nuclear adiabatic demagnetization refrigerator is at least less than 12 mK. This is further corroborated by comparing the Hall resistance of several reentrant IQH states. As one can see from Fig. S4, the data collected in the nuclear demagnetization refrigerator have better quality, which again suggests that it reaches electron temperature lower than 12 mK (the value in our dilution refrigerator). On the other hand, it is quite likely that the conventional sample holder in our refrigerator is less efficient in cooling of the electron than the $^3$He immersion cell used in Ref.[23], which achieved an electron temperature of 4 mK.

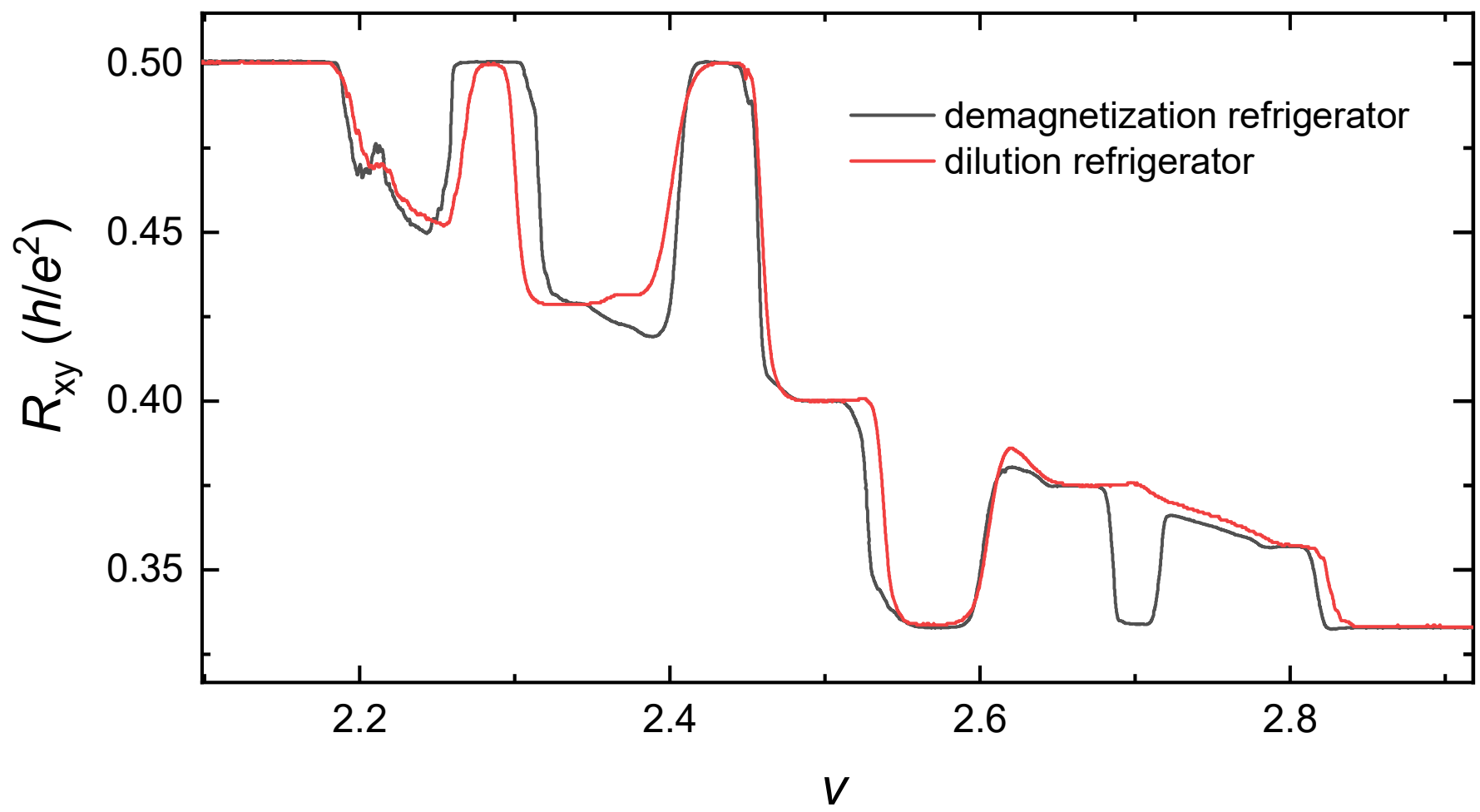

**Figure S4.** The Hall resistance of several reentrant IQH states in the second Landau level measured in the nuclear adiabatic demagnetization refrigerator and the dilution refrigerator.